\documentclass[conference]{IEEEtran}
\IEEEoverridecommandlockouts
\pdfoutput=1
\usepackage{cite}
\usepackage{amsmath,amssymb,amsfonts}
\usepackage{algorithmic}
\usepackage{graphicx}
\usepackage{textcomp}
\usepackage{xcolor}

\usepackage{booktabs}
\usepackage{makecell}
\usepackage{multirow}

\def\BibTeX{{\rm B\kern-.05em{\sc i\kern-.025em b}\kern-.08em
    T\kern-.1667em\lower.7ex\hbox{E}\kern-.125emX}}
\begin{document}


\newcommand{\loremipsum}[1]{\noindent{{\color{gray}{\sc \bf LOREMIPSUM:}   #1}}}
\newcommand{\rev}[1]{\noindent{{\color{red!80!black}{\sc \bf REV:} \bf   #1}}}
\newcommand{\ahmad}[1]{\noindent{{\color{green!80!black}{\sc \bf AMK:} \bf   #1}}}
\newcommand{\naw}[1]{\noindent{{\color{olive!80!black}{\sc \bf NSS:} \bf   #1}}}
\newcommand{\woong}[1]{\noindent{{\color{blue!80!black}{\sc \bf WS:} \bf   #1}}}
\newcommand{\feiyi}[1]{\noindent{{\color{red!80!black}{\sc \bf FW:} \bf    #1}}}

\title{
Profiling and Modeling of Power Characteristics of Leadership-Scale HPC System Workloads

\thanks{This work was supported by, and used the resources of, the Oak Ridge Leadership Computing Facility, located in the National Center for Computational Sciences at ORNL, which is managed by UT Battelle, LLC for the U.S. DOE (under the contract No. DE-AC05-00OR22725).
The US government retains and the publisher, by accepting the article for publication, acknowledges that the US government retains a nonexclusive, paid-up, irrevocable, worldwide license to publish or reproduce the published form of this manuscript, or allow others to do so, for US government purposes. DOE will provide public access to these results of federally sponsored research in accordance with the DOE Public Access Plan (http://energy.gov/downloads/doe-public-access-plan).}

}

\author{\IEEEauthorblockN{Ahmad Maroof Karimi}
\IEEEauthorblockA{\textit{Oak Ridge National Laboratory}\\
Oak Ridge, TN, USA \\
karimiahmad@ornl.gov}
\and
\IEEEauthorblockN{Naw Safrin Sattar}
\IEEEauthorblockA{\textit{Oak Ridge National Laboratory}\\
Oak Ridge, TN, USA \\
sattarn@ornl.gov}
\and
\IEEEauthorblockN{Woong Shin}
\IEEEauthorblockA{\textit{Oak Ridge National Laboratory}\\
Oak Ridge, TN, USA \\
shinw@ornl.gov}
\and
\IEEEauthorblockN{Feiyi Wang}
\IEEEauthorblockA{\textit{Oak Ridge National Laboratory}\\
Oak Ridge, TN, USA \\
fwang2@ornl.gov}
}

\maketitle

\begin{abstract}

In the exascale era in which application behavior has large power \& energy footprints, per-application job-level awareness of such impression is crucial in taking steps towards achieving efficiency goals beyond performance, such as energy efficiency, and sustainability.

To achieve these goals, we have developed a novel low-latency job power profiling machine learning pipeline that can group job-level power profiles based on their shapes as they complete.  This pipeline leverages a comprehensive feature extraction and clustering pipeline powered by a generative adversarial network (GAN) model to handle the feature-rich time series of job-level power measurements. The output is then used to train a classification model that can predict whether an incoming job power profile is similar to a known group of profiles or is completely new.
With extensive evaluations, we demonstrate the effectiveness of each component in our pipeline.  Also, we provide a preliminary analysis of the resulting clusters that depict the power profile landscape of the Summit supercomputer from more than 60K jobs sampled from the year 2021.



\end{abstract}

\IEEEpeerreviewmaketitle

\begin{IEEEkeywords}
HPC Job Power Consumption, 
Job Power Profile Clustering, 
Job Power Profile Prediction, 
Time series Analysis, 
GAN
\end{IEEEkeywords}

\section{Introduction}
\label{sec:intro}

With the extreme energy footprint of modern exascale-era supercomputers, large-scale HPC sites need to take energy, a quantitative physical metric, into account when setting efficiency goals on top of their performance goals improving their operations\cite{ref:4pillar:2014}. In such a process, it is crucial to understand the behavior of large-scale HPC applications since the applications themselves define the power consumption of an HPC system by driving the power-hungry compute components. Algorithmic behavior at the code level, at scale, translates directly into a power envelope of a few tens of megawatts. This has a profound impact on how the underlying HPC data center responds in terms of power distribution, cooling, and long-term reliability \cite{shin2021revealing}. Being able to profile applications by systematically putting power and energy measurements into a sub-cluster job-level, application context can benefit supercomputer operations by enabling cross-layer use cases leading towards energy-aware resource utilization at the system, software, and building infrastructure levels.

However, systematically acquiring and interpreting such power profiles is a non-trivial task due to the difficulty in processing features and modeling on the high-volume data streamed from these large systems. While making use of fine-grained power profiles per application have been used for the analysis during the lifetime of the supercomputer~\cite{naghshnejad2018adaptive,galleguillos2018data,wyatt2018prionn,mckenna2016machine,emeras2017evalix1,sirbu2016power,matsunaga2010use,galleguillos2020accasim,klusavcek2020alea}, such profile information has been limited to a non-behavioral granular level or for long-term retrospective analysis. The hidden behavioral insights from these data streams are either locked up until late or are discarded without use. This necessitates an automated method that can handle the volume and velocity of these data streams, tapping into the feature-rich dynamic behavior recorded in the form of timeseries. The absence of meaningful ground truth labels that represent the characteristics of the power timeseries of HPC jobs makes it additionally challenging to quantify and classify the workloads for system-wide profiling, monitoring and classification, and none of the previous work has done it.

In response, we have developed a machine learning based job power profile processing pipeline that can classify the entirety of jobs executed on an HPC system based on the characteristics of the power consumption pattern. Targeted for pre-exascale and exascale supercomputers such as Summit~\cite{summit} and Frontier~\cite{frontier}, the pipeline operates on streams of high-resolution high-volume out-of-band power and energy measurements per component data from these systems~\cite{shin2021revealing, Thaler:2020, openbmc_event}, grouping 10-second interval job-level timeseries power profiles as they are ingested. These profiles are
grouped based on their temporal characteristics, such as swings, magnitude, average, and maximum in-power consumption, helping profile and assess the landscape of power consumption of the workloads and, thus, the complete HPC system.

To achieve this, we have developed a comprehensive feature extraction pipeline that extracts 186 distinctive features from timeseries data and subsequently processes it to train an open-set neural network classifier capable of categorizing jobs into one of the known classes or unknown/unseen group. Since the raw data is unlabeled, we develop a clustering model to group historical workloads exhibiting similar power behavior, and this helps us generate labels with meaningful contexts, such as different severity of compute-intensive and non-compute-intensive workloads. These labels eventually assist in giving context to the classification of the neural network classifier. Overall, the clustering model generates contextualized classes based on historical data, and the classifier model enables us to predict the label of the new jobs as they conclude. 
Given the high-dimensional complexity of the data, we employ a generative adversarial network (GAN)~\cite{tadgan, clustergan, gan8594983} to simplify the data while enhancing clustering precision.  Further, by periodically conducting iterative offline processing and training with continually expanding historical data, the classifier enables classification of evolving workloads by incorporating significant new pattern as known job patterns. 
This adaptive process ensures the pipeline remains responsive and effective in dynamic computing environments.
With this pipeline, our contributions are as follows:


\textbf{Design and implementation of a job power profile clustering and classification pipeline for large-scale HPC systems:}
We present a job power profile clustering and classification pipeline, purpose-built for monitoring large-scale HPC facilities experiencing new workload patterns. This solution tackles the unique challenges posed by high-volume, feature-rich, variable-length timeseries data originating from an ever-evolving blend of workloads. 
The pipeline efficiently classifies HPC job power profiles based on their temporal characteristics, empowering HPC facilities with energy-aware decision-making capabilities and optimizing resource usage in dynamic computing environments.

\textbf{Feature extraction and unsupervised learning on HPC job profile data to provide contextualized labels:}

We also present the implementation details of the feature extraction and clustering of timeseries-based power profiles engineered to satisfy downstream applications such as classification and provide a method to classify newly completed jobs using low latency inference classification.  In particular, we identify and report the challenges we have found while employing the pipeline and ensuring the outcome is valid and has a meaningful context.

\textbf{Open-set classification for handling unknown power profiles:}
One of the challenges of applying classification to real-world data is that incoming data is always evolving and the classification inference model may encounter new data that the model has not seen during the training phase. The open-set classification model reliably classifies data from known classes as well as unknown data.

\textbf{Insights on grouping HPC jobs based on their power profile:}
We present an analysis of what our clustering pipeline has found from the system-wide data we used, which was from the entire year of 2021, that reveals the power utilization landscape of the HPC system. We describe the correctness, usefulness, and areas of improvement, helping ourselves establish a firm foundation for long-term operations and maintenance of the system.

\section{Background and Motivation}


\subsection{Job Power Profiling, Classification, and its Usage}
\label{subsec:use-cases}
In the context of (pre-)exascale HPC systems, obtaining contextualized job-level sub-cluster power consumption profiles 
can help gain insights into
improving the energy efficiency of HPC systems and their data centers.  In the following, we describe the opportunities that motivated us.


\subsubsection*{\textbf{Continuous monitoring of application behavior and compute-node components}}
Applications themselves define the high-level energy usage of an HPC user facility. Any unusual change in their behavior will be reflected in the power pattern that they exhibit. If there is a significant change in the application behavior, job level profiling and classification will classify the jobs into a class differently than it used to classify earlier or into an unknown group. It can also enable the operations teams to quickly identify the sub-optimal 
conditions.
Correlating application behavior with power consumption opens avenues for energy awareness in HPC in the face of numerous challenges lying in post-exascale HPC~\cite{amd-supercomputer-gains:2023}

\subsubsection*{\textbf{Long-term performance analysis and energy driven design and procurement}}
Evaluation of long-term behavior of various HPC component's performance and their degradation.
At the same time, it can help us observe the long-term evolution of HPC applications and monitor any 
trends. Observations from general trends can feed into better accounting of building infrastructure, system hardware design, and acquisition, ultimately leading to more energy-efficient HPC systems. This requires a comprehensive assessment of application power
profiles throughout the lifetime of a system.



\subsubsection*{\textbf{Power and energy usage prediction for intelligent resource usage}}
If implemented with real-time job classification and prediction, power profiling can serve as a foundation for intelligent resource usage. For example, one HPC energy-efficiency-driven use-case is optimizing cooling operations for HPC systems by informing cooling systems to make better staging and de-staging decisions for cooling resources such as chillers and cooling towers. Also, for resource managers, online job power profile classification provides the potential for better allocation and can be used to leverage cheaper computing power in terms of potential energy usage driven by application behavior.

\subsection{Challenges}
\label{subsec:challenges}

The system-wide power profiling and classification of large HPC machines come with several challenges. To study, analyze, and model the data from multiple sources, such as system-level sensor data and application-level jobs logs, should be synchronized and merged to make it suited for application-level power consumption analysis. We briefly describe these challenges as follows:


\subsubsection*{\textbf{Unseen data from unknown distribution}}
\label{subsubsec:unseen_data}
One of the challenges of the real-world project is that the system is always evolving. Jobs generating new patterns can be executed on HPCs without any prior information. The machine learning models are reliable in handling the data from seen distribution or known classes, however, they perform poorly when exposed to unknown classes. We developed an open-set classifier to handle unknown data.

\subsubsection*{\textbf{Representation of features}}
\label{subsubsec:representation}
Representation of timeseries by features extracted from the jobs is crucial as all the subsequent work depends on the features extracted from the timeseries. We meticulously identified the features that can effectively capture the behavior of the jobs such as swings, plateaus, slopes, and magnitude.


\subsubsection*{\textbf{Variable non-uniform length of profiles}}
\label{subsubsec:low_latency}
In this work, we converted the raw data into job-level power profile timeseries. The timeseries data provides high-resolution behavior of the workloads. However, one of the challenges of the timeseries is that it depends on the length of the job and every job has a different runtime and thus a different timeseries length. Thus, extracting information/features from the timeseries data of the fixed-length vector of all jobs become critical as most of the data-driven analysis and modeling requires fixed-length representation.

\section{Design}
\label{sec:design}

\subsection{Design Goals}

In developing our system-wide HPC workload profiling and monitoring pipeline, we have identified three core design goals.
These design goals enable us to generate contextualized labels for the power profiles in the absence of any ground truth labels and align with our overarching requirement of continuously providing a systematic understanding of power \& energy. 


\subsubsection*{\textbf{Cluster HPC Job Power Profile}}
One of our primary design objectives is to analyze and study HPC workload power characteristics. This requires us to effectively contextualize HPC job power profiles so that we can identify the most dominant and critical profiles exhibited by HPC workloads. The profiles consist of job-level power utilization timeseries data that varies in magnitude and length based on the scale and duration of the job. Our focus is on employing methods that can extract robust features capable of identifying similarities despite the inherent complexity and variability of these power profiles. Specifically, we strive to ensure that the features accurately reflect requirements in recognizing fluctuations, swings, and patterns in power consumption resulting from dynamic HPC application behavior. Clustering provides labels that help build a classification model that continuously monitors the HPC system. 

\subsubsection*{\textbf{Low-latency Classification and Recognition of New Data}}
The goal is continuously monitoring, i.e., classifying incoming
streaming job profiles as they are completed. It is crucial to quickly identify and handle unknown new profiles (unknown class or distribution) and the existing known ones in a reliable manner. Our focus is on designing methods that can correctly group known job profiles while differentiating them from unknown or newer patterns. Additionally, we need to find the class if the job belongs to a known class. Classification has to be computationally inexpensive so we can immediately infer the class of the incoming data point. 

Clustering is a computationally expensive job in which the entire process in our case may take over a day to render contextualized labels on historical data.  Our classification model is an inference model that provides the labels instantly for the new and recently completed jobs and enables us to monitor the HPC system continuously.


\subsubsection*{\textbf{Iterative workflow Management}}
To address the difficulty of dealing with ever-evolving applications and workloads, the monitoring and classification system are exposed to newer patterns or classes that have not been previously encountered, our third objective is to create a method that can effectively maintain the accuracy of our clustering results over an extended period and not predict the classes of datasets that do not belong to already known classes. When new power profile patterns either become critical or frequent enough and the number of unknown data points becomes increasingly large, then we need to incorporate these new patterns into the pipeline first by assigning them a new contextualized label and then training the classifier with the dataset including the new labels. For this kind of task, we periodically run the clustering module to identify this new pattern by capturing it in one of the clusters, and then, we assign it a new label to incorporate in the classification pipeline. Here, we manually visualize the clustering results to ensure that the data points in the cluster are homogeneous and make sense before making the data points of that cluster a new class. While it is ideal if the handling of unknowns is fully automated, for the absence of ground truth labels and for reliability, we envision human involvement in the process of introducing new classes in the pipeline.  
\subsection{End-to-end Pipeline}
\label{subsec:end-to-end}
\begin{figure}[t]
  \centering
  \includegraphics[width=0.95\linewidth]{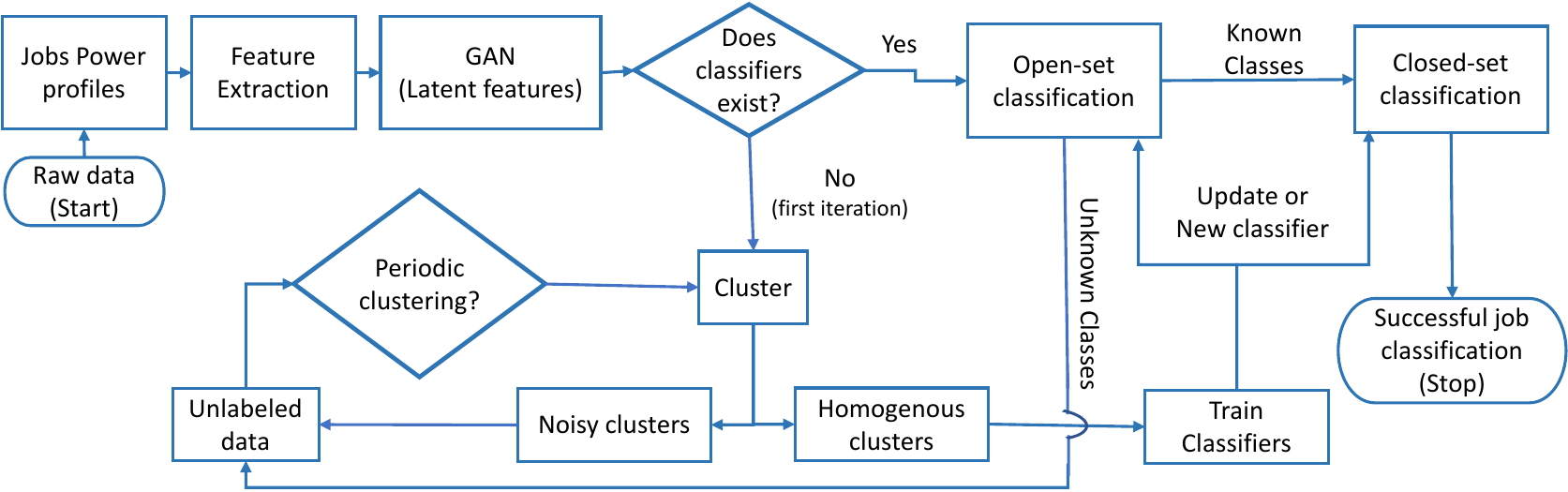}
  \caption{End-to-end flow diagram of the data-driven analysis and classification modeling pipeline. }
  \label{figs:end2end}
\end{figure}

We employ data-driven techniques to achieve our objective of profiling the power characteristics of HPC workloads, identifying crucial patterns, and their classification. 
The major steps of the pipeline are data processing, feature extraction, clustering, open-set, and closed-set classification, and maintaining iterative workflow (Figure~\ref{figs:end2end}).


\if
\ahmad{
\subsubsection*{\textbf{Data ingest and data processing:}}
Our system relies on a separate data engineering and data distribution system that is responsible for collecting telemetry data from heterogeneous sources into a single location and then distributing it to downstream users as mentioned in Section~\ref{subsec:oda_arch}.  Among many telemetry data stream “topics”, our job power profiling mainly focuses on two data sources: 1) the job allocation data which comprises job allocation events \& node-level state change, and 2) out-of-band high-resolution per-node, per-component (i.e., CPUs, GPUs) telemetry stream from the entire HPC cluster as seen in Figure~\ref{figs:end-to-end}-a).

Following data ingestion, the data undergoes extensive processing to suit downstream needs (Figure~\ref{figs:end-to-end}-b)). We focus on contextualizing power measurement data into a job allocation context by merging two data streams based on their timestamps and hostnames. This process includes data aggregation for timestamp alignment and pre-computation of variances and quantiles across various dimensions such as time, space, and allocated nodes. Pre-computations are performed at both job and component levels.  Due to the high-volume nature of the raw data ingest, we employ a dedicated data analytics cluster based on Apache Spark~\cite{spark} and leverage its structured streaming API~\cite{spark-structured-streaming} as a unified way of implementing data processing code for both new data in the streaming context (Figure~\ref{figs:end-to-end}-b) upper) and historical data (Figure~\ref{figs:end-to-end}-b) bottom) with a single copy of code.
}
\fi

The objective of the data processing module is to generate a job-level power profile or power timeseries for each workload using telemetry data and the scheduler logs (shown in Table~\ref{tab:data_spec}). It takes the raw data as input to generate the job-wise processed data that can be used by subsequent modules such as feature extraction and clustering methods. 
During the feature extraction step, we calculate relevant features that can capture swings, slopes, and magnitudes in the job power timeseries. To capture different aspects of power characteristics, we computed 186 features. For efficient clustering, we reduce the dimensionality of the data such that the reduced feature set results in a dense dataset but still contains all the useful information. We use a generative adversarial network (GAN) variant to reduce the dimensionality to 10 features. These features are critical for the computationally efficient clustering of jobs into different groups based on their power characteristic. The power profile data inherently do not provide job groupings or ground truth labels based on power characteristics. Therefore, to monitor and analyze the power characteristics, we cluster the jobs into multiple classes based on the compute intensity and patterns of jobs. The clustering technique allows us to group jobs with similar profiles, thus providing contextualized labels and enabling us to get insights into the nature of the workloads.

\begin{table}
    \caption{Datasets Description(1 Jan 2021 to 31 Dec 2021)}
  \label{tab:data_spec}
  \footnotesize
  \setlength\tabcolsep{4pt}  
  \begin{tabular}{cm{14mm}m{10mm}m{8mm}m{32mm}}
    \toprule
      id & Name & Resolution & Rows  & Description \\
    \midrule

      (a) & Job scheduler 
      & per-job & 1.6M  & Project, allocation param. submit, start \& end time\\ 
      (b) & Per-node job scheduler 
      & per-job  & 9GB & Per-node job allocation history end of job statistics \\ 
      
      (c) & Power telemetry
      & 1 sec & 268B & System-level per-node, per-component power \\ 
      (d) & Job-level processed data& 10 sec & 201M & Job level power data aggregated over all the compute-nodes \\
    \bottomrule
  \end{tabular}
\end{table}

After we find the relevant clusters, we train a classifier based on the labels generated in the clustering step. As mentioned earlier, the classifier provides a low-latency classification of the new workloads for continuous monitoring. The classification step consists of two parts, i.e., closed-set and open-set classification models~\cite{Toward_Open_Set_Recog,miller2021class}. The closed-set
classification is a traditional neural network classifier that classifies the input data into one of the known classes. The classifier ensures that if the incoming data point is from the known classes, then the power characteristic of an incoming job can
be classified into one of the known classes on which the model
was trained. The open-set classification is a task that identifies
whether the incoming job is from one of the known classes
or distribution or whether it is from an unknown class. This
is critical because often, the HPC system experiences a job
exhibiting new patterns, and if we use only the closed-set
classifier, it will always classify the incoming job into one of
the known groups, although the new data point may not belong to any existing classes. 

The end-to-end pipeline is designed such that it can be iteratively updated. The iterative workflow provides an opportunity to allow the pipeline to adapt to the changing HPC workload power pattern. If there are new or existing applications that generate a new power pattern different from all the known
classes due to changes in the underlying workloads. In that
case, the pipeline should be updated to handle the
changes in HPC workloads. During the iterative update, we
identify new classes and update the open-set and closed-set
classifiers to adapt to a new set of known classes for prediction.

\section{Implementation Details}
\label{sec:details}
In this section, we present the implementation details of the different parts of the pipeline.
The end-to-end pipeline contains a comprehensive set of data-processing and machine-learning modules to manage large-scale data, and build clustering methods and classification models.
\subsection{Data Processing}
\label{subsec:details:processing}

The data-processing step takes the raw telemetry data captured at the system-level and the job information from the HPC scheduler logs. The system-level power telemetry data are captured at a 1-Hz frequency on every compute node, but
this data does not include any job details.
Thus, for every job, we get the job information such as compute nodes on which the job was executed and the duration of the job execution from scheduler logs, and then, using the telemetry data, we extract power values of those compute nodes on which the job was executed and for the duration for which the job was executed. The aggregated power values of these compute nodes for the job duration give the power profile of each job. 

Table~\ref{tab:data_spec} provides a brief description of the data. Dataset (a) and (b) are job scheduler logs that primarily provide job information like \texttt{job\_id}, \texttt{start-time}, \texttt{end\_time}, \texttt{hostname} and other important fields. 
Dataset (c) contains the telemetry data from \texttt{hostname} or compute nodes, it contains the compute node's power values for each component of the compute node including input power at 1-second frequency. 
The dataset (d) in Table~\ref{tab:data_spec}~which is the output of the data processing pipeline is generated by combining datasets from (a) to (c). To generate the output data, we first reduce the frequency of the source telemetry data from 1 sec. to 10 sec. by taking the mean value of the input power for every 10-second window for each compute node. This also helps us eliminate the issue of missing values in 1-Hz dataset. Then, for every job, we find out the compute nodes on which the job was executed and take the mean value across all the nodes of the job. This per-node normalized data is output data. The per-node normalization of jobs power timeseries enables us to compare different jobs running on different numbers of nodes.   
Figure~\ref{fig:classes_groups} represents the per-node normalized power profile of some of the typical HPC jobs.

\subsection{Feature Extraction}
\label{subsec:details:feature}

\begin{figure}[t]
  \centering
  \includegraphics[width=0.95\linewidth]{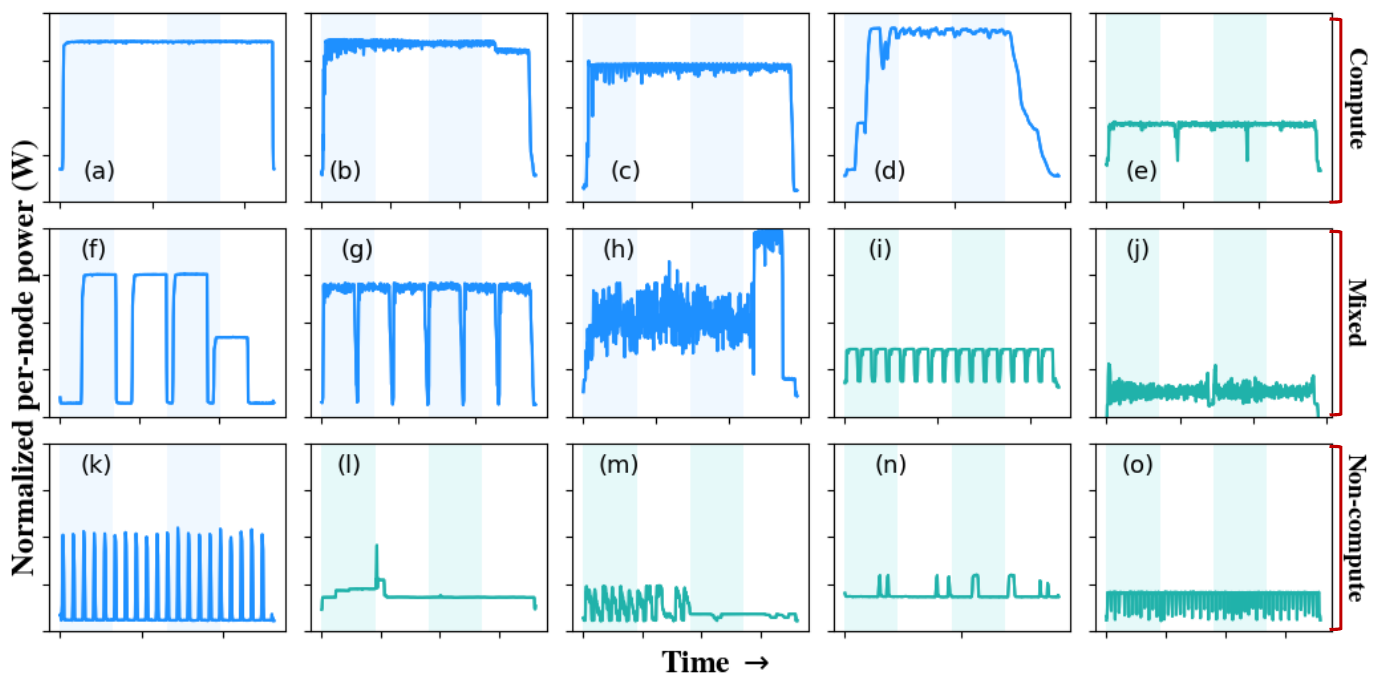}
    \caption{Timeseries of typical HPC workloads. The subfigures help visualize that the features calculated in Table \ref{tab:data_feature_spec} can distinguish between different power profiles. Background shades in each subplot correspond to 4 bins of timeseries.}
  \label{fig:classes_groups}
\end{figure}

The relevant characteristics of data (power profiles) are calculated so that the data points with apparently similar attributes result in similar values for computed features. 
As mentioned earlier in Section~\ref{subsec:challenges}, power profile timeseries having varying lengths are characteristic of jobs that depend on the time length for which the HPC jobs ran. Varying timeseries are not ideally suited for machine learning algorithms as machine learning methods require all data points to have equal dimensions/lengths. Thus, feature extraction has an added advantage, particularly in timeseries analysis and clustering, that the length of the extracted features vector will be the same for all power profiles and we normalized the feature values to make it independent of the duration/length of the HPC jobs. 
From the HPC power facility's perspective, the three essential characteristics with the highest potential to adversely affect are the frequency of power swings, slope, and the range of magnitude of those swings. While calculating features, we focus on these three properties of power profiles for feature extraction. We also preserve the notion of temporal relevance when extracting the features by dividing the timeseries into four bins of equal time length, as highlighted by four regions in each subplot of Figure~\ref{fig:classes_groups}. 

The partitioning of timeseries into four bins allows us to maintain the partial temporal characteristics of power consumption signals. The binning of the timeseries enables us to calculate the features for each bin in addition to the aggregate values over the complete timeseries. Table \ref{tab:data_feature_spec} shows the summary of features that were extracted. For brevity, we will only describe a subset of features, but other features follow similar patterns. Feature \texttt{$[\text{*}]$\_mean\_input\_power} refers to the mean input power of four bins and is represented as \texttt{$\text{1}$\_mean\_input\_power}, \texttt{$\text{2}$\_mean\_input\_power}, \texttt{$\text{3}$\_mean\_input\_power}, and 
\texttt{$\text{4}$\_mean\_input\_power}. 
Similarly, feature \texttt{$[\text{*}]$\_sfq[p/n]\_$[\text{\#}]$\_$[\text{\#}]$} refers to rising and falling swings of all four bins and of magnitudes in the range of \texttt{50W-100W}, \texttt{100-200W}, \texttt{300-400W}, \texttt{400-500W}, \texttt{500-700W}, \texttt{700-1000W}, \texttt{1000-1500W}, \texttt{1500-2000W}, \texttt{2000-}\texttt{3000W}. 
Three of the sample features are: 1) \texttt{$\text{1}$\_sfqp\_$\text{50}$\_$\text{100}$} is a feature for bin one that counts rising swings in the range of \texttt{50W} to \texttt{100W}, 2) \texttt{$\text{1}$\_sfqn\_$\text{50}$\_$\text{100}$} is also a feature for bin one but counts falling swings in the range of \texttt{50W} to \texttt{100W}, and  3) \texttt{$\text{4}$\_sfqp\_$\text{1500}$\_$\text{2000}$} is a feature for fourth bin that counts rising swings in the range of \texttt{1500W} to \texttt{2000W}. The \texttt{length} feature is calculated to normalize other features by dividing them by the length of the timeseries. This is because features having swing counts will have a larger value for longer timeseries, and a similar power profile will have a smaller swing count if the timeseries are shorter.
All of the features we calculated help us profile timeseries based on the HPC facility team recommendation.
\begin{table}
    \caption{Summarized List of 186 features calculated from timeseries of each of the HPC workloads}
  \label{tab:data_feature_spec}
  \begin{tabular}{ll}
    \toprule
      Feature & Description \\
    \midrule
    {[}*{]} \_mean\_input\_power & Mean input power\\
    {[}*{]} \_median\_input\_power & Median input power\\
    {[}*{]} \_sfqp\_{[}25{]}\_{[}50{]} & Count of rising swings of 25 to 50 W\\
    {[}*{]} \_sfqn\_{[}25{]}\_{[}50{]} & Count of falling swings of 25 to 50 W\\
    \multirow{1}{*}{{[}*{]} \_sfq{[}p/n{]}\_{[}\#{]}\_{[}\#{]}}  
                           &  Similarly, rising and falling swings \\
                           & 50-100W, 100-200W, 300-400W,\\ 
                                        &  400-500W, 500-700W, 700-1000W, \\ 
                                        & 1000-1500W, 1500-2000W, 2000-3000W \\
    {[}*{]} \_sfq2p\_{[}25{]}\_{[}50{]} & Rising swings 25-50 W over lag of 2\\
    {[}*{]} \_sfq2n\_{[}25{]}\_{[}50{]} & Falling swings 25-50 W lag of 2\\
    \multirow{1}{*}{{[}*{]} \_sfq2{[}p/n{]}\_{[}\#{]}\_{[}\#{]}} 
     & Similarly, rising and falling swings\\
      & 50-100W, 100-200W, 300-400W,\\
  &400-500W, 500-700W, 700-1000W,\\
                                        &1000-1500W, 1500-2000W, 2000-3000W\\ 
                                        &over lag of 2 timeperiod\\    
    mean\_power & Mean value of whole timeseries \\
    length & Length of timeseries\\
  \bottomrule
\end{tabular}
\end{table}
\vspace{-2pt}
\subsection{{Generative Adversarial Network (GAN)}}
\label{subsec:details:gan}

GANs have recently been deployed in various unsupervised learning and timeseries analysis works~\cite{tadgan, clustergan, gan8594983}. GAN models provide a robust method to generate low-dimensional latent space $(\mathbb{R}_{z})$ representation of the high-dimensional $(\mathbb{R}_{x})$ real data and also captures the distribution of high-dimensional data. 
The GAN model we have used benefits from the techniques of modified GAN models because they have certain advantages over basic GAN model architecture~\cite{tadgan, clustergan, infogan, wasserstein}. Our GAN model is inspired by TadGAN model presented in~\cite{tadgan}. 


\begin{figure}[t]
  \centering
  \includegraphics[width=0.8\linewidth]{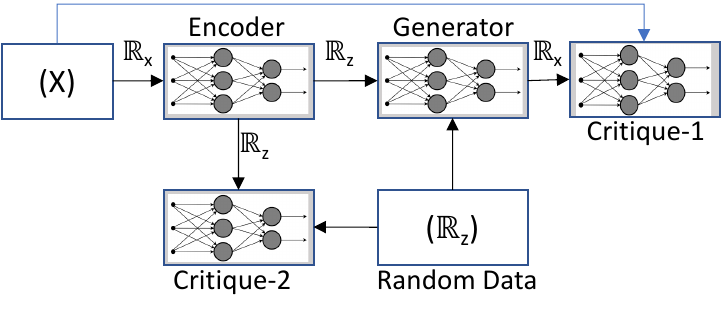}
  \caption{GAN model for latent feature generation. It takes input of dimension $\mathbb{R}_{x}$ and generate the latent feature of lower dimension $\mathbb{R}_{z}$.
  Once trained, the model can generate latent features for new data points.}
  \label{figs:gan_arch}
\end{figure}
The model in  Figure~\ref{figs:gan_arch} shows the architecture of the GAN model we used to generate the latent features. A typical GAN model has a generator and a discriminator(or critique) and usually it takes a random fixed-length vector to generate the real looking data (generated data) and then critique is fed both the real and generated data and it tries to classify them correctly. In the GAN model that we implemented, in addition to Generator $(G)$ and Critique-1 ($\texttt{C}_{1}$), the model also has Encoder $(E)$ and another Critique-2 ($\texttt{C}_{2}$). Instead of using the random data of fixed length to generate the ouptut, in this method actual data (\texttt{X}) is fed first to Encoder to generate a feature set in the latent feature space. The Generator takes the output of the Encoder and maps the data from latent space$(\mathbb{R}_{z})$ to high-dimensional space($\mathbb{R}_{x}$) of real data. Critique ($\texttt{C}_{1}$), called Discriminator in the basic GAN model, uses the Wasserstein method to calculate the loss in classifying real and generated data. 
Traditional GAN model uses BCE loss function which is described in Equation~\ref{eqn:modified_bce} as: 
\begin{equation}
\mathcal{L} = \underset{\texttt{C}_{1}}{min}~\underset{gen.}{max} -[\mathbb{E}(\log (\texttt{C}_{1}(x)))+\mathbb{E}(1-\log~(\texttt{C}_{1}(G(z)))) ]   \label{eqn:modified_bce} 
\end{equation}
The issue with Equation~\ref{eqn:modified_bce} is that the loss value ($\mathcal{L}$) ranges only between 0 and 1 and thus has a vanishing gradient problem. This leads to mode collapse or in other words, vanishing gradient early stops the generator function training so that the generator can not learn to generate all the patterns in the data. 

\begin{figure}[t]
  \centering
  \includegraphics[width=0.8\linewidth]{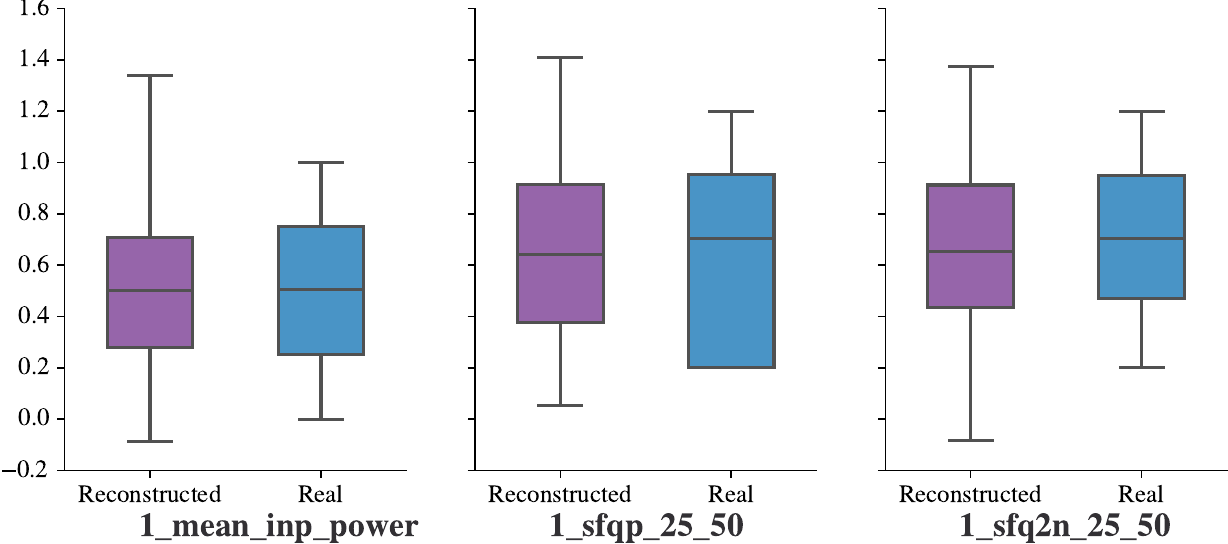}
  \caption{Data distribution of three reconstructed and real features showing that the reconstructed distribution is similar to the real distribution. Reconstruction of original features confirm that the low-dimension latent feature have necessary information of the data-points as in the high-dimension space.}
  \label{figs:gan_distribution}
\end{figure}

Wasserstein\cite{wasserstein} loss function in Equation \ref{eqn:Wloss} does not have a bounding range of 0 \& 1 and can continue to grow beyond 1. Thus, the gradient of the loss function never approaches zero and the GAN models are not prone to the vanishing gradient problem.
\begin{equation}
    \mathcal{L} = \underset{gen.}{min}~\underset{\texttt{C}_{1}}{max}~\mathbb{E}(\texttt{C}_{1}(x))~-~\mathbb{E}(\texttt{C}_{1}(G(x)))
 \label{eqn:Wloss}
\end{equation}
Wasserstein loss helps the model overcome the mode collapse problem and enables the GAN to learn all different patterns in the data. 
Critique ($\texttt{C}_{2}$) discriminates between random data and the encoded signal. It enables the Encoder to learn to generate the distribution of latent vectors, which recreate the real data with a distribution similar to the original distribution. The reconstruction-based method is useful in generating the dataset using a latent feature set. It also ensures that the latent features of lower-
dimensional data ($\mathbb{R}_{z}$) represent the information in higher-
dimensional, i.e., real data space. We observe the results in Figure~\ref{figs:gan_distribution}, and we validate that the distribution of the generated dataset is similar to the real or actual data. 
This ensures that the latent features generated by the Encoder are true representations of the actual data, thus Encoder can be used to generate the latent features of incoming data as well.

The input to the GAN model, shown in Figure \ref{figs:gan_arch}, is the dataset (\texttt{X}) of extracted features vectors ($\mathbb{R}_x$) of size 186 and outputs latent vectors ($\mathbb{R}_z$) of size 10. 
The encoder model and the generator model both have two linear layers and a batch normalization layer between the two linear layers. The input and output dimensions of the two encoder linear layers are $186\times40$ and $40\times10$ and the dimensions of the two generator layers are $10\times128$, and $128\times186$. The activation function is \texttt{ReLU}. Critique's ($C_1$) three linear layers have input and output dimensions of $10\times100$, $100\times10$, and $10\times1$. Critique ($C_2$) has one linear layer of input and output dimensions $10\times1$. 
When the model is successfully trained we use the Encoder to generate the latent features from the real data for clustering. It also ensures that once the model is trained, every job will have deterministic representation in the latent vector space. 
\subsection{Clustering  Based on Workloads Power Pattern}
\label{subsec:details-clustering}
We want to identify and group workloads that exhibit the most prevailing, essential, and critical patterns. 
The groupings enable us to identify the jobs based on the compute intensity depending on their class thus enabling us to have contextualized labels. We apply the density-based spatial clustering DBSCAN~\cite{dbscanMethod}~method onto the latent features generated by GAN to cluster jobs into different classes. It works on the principle that the clusters are formed by dense regions where the data points are concentrated in the feature space and are separated by sparse regions of no or light density of data points. The data points that do not belong to any cluster are labeled noise data. 

\begin{figure}[t]
  \centering

  \includegraphics[width=\linewidth]{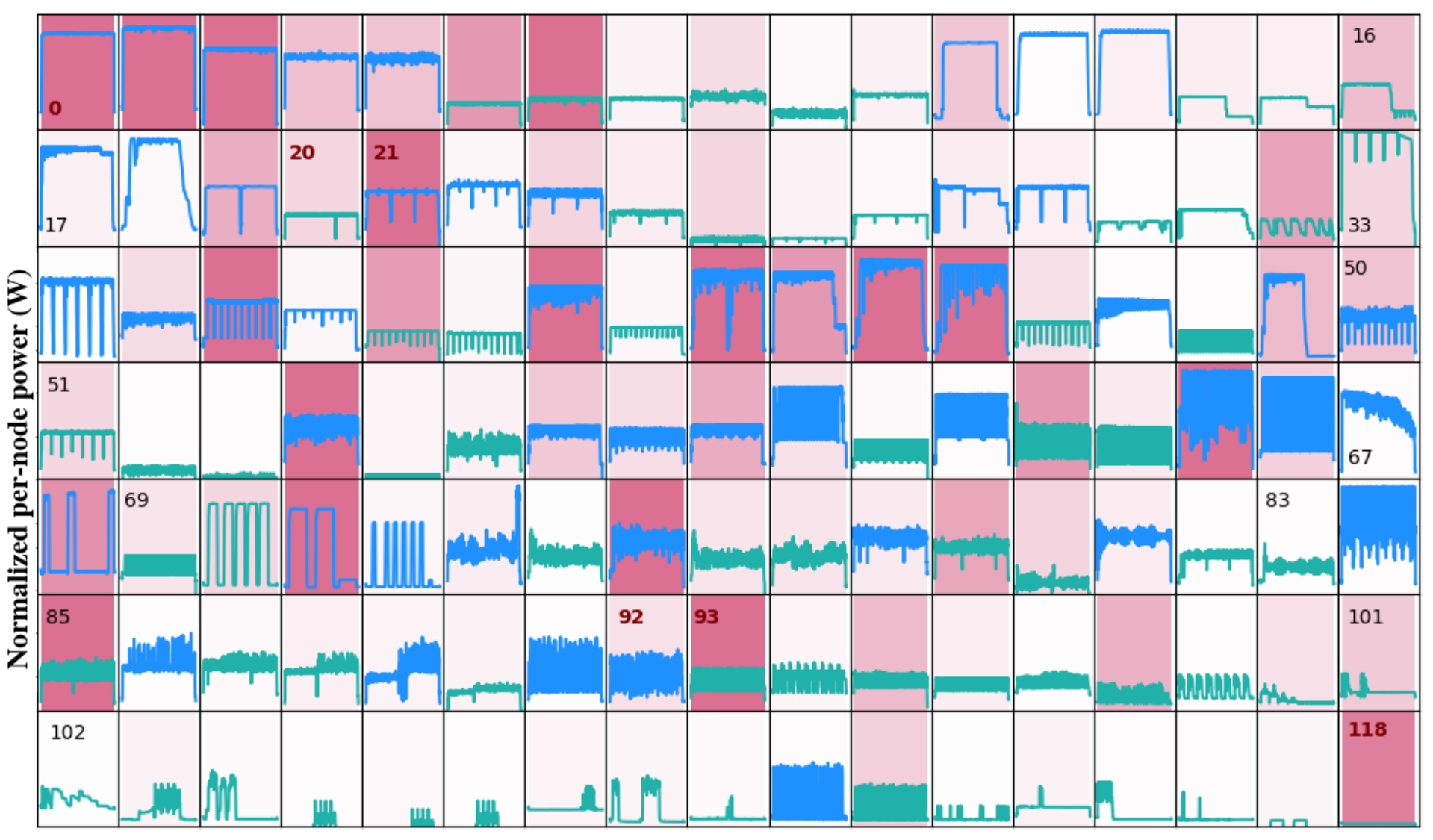}
    \caption{Groupings of power profiles based on the power utilization patterns exhibited by jobs. The different classes can be grouped into high-level classification into three types. Classes from 0-20 are compute intensive, 21-92 are mixed type, and 93-118 are non-compute intensive jobs. Tiles with blue curve have high power consumption and plots with green curve have low power consumption.  Intensity of the background color of each tile represents the density of each pattern among the entire population.}
  \label{fig:all_class}
\end{figure}

Overall, we grouped about $60,000$ workloads into $119$ classes.  
Figure~\ref{fig:all_class}~shows the power profiles of representative jobs of the $119$ classes. Each class has unique patterns and maintains relative similarity and dissimilarity with other classes. 
Based on the power consumption of jobs represented by the patterns, we classified the jobs into three groups: compute intensive, non-compute intensive, and mixed operation jobs. The classification analysis is shown in Table~\ref{tab:intensity}. Based on the magnitude of the power consumption, which is dependent on the component used (i.e., CPU, GPU, and certain GPU kernel), we further classify the jobs into High and Low based on the high power or low power for most of the duration during their runtime.

\subsection{Classification Model}
\label{sec:details:openset}
Clustering is a computationally expensive task that takes a long time to cluster the historical data points and generate meaningful contextualized labels. However, clustering cannot satisfy our low-latency requirements for handling new data for monitoring the HPC system. In this effort, we build a classifier that can be trained based on the power data and class labels identified by the clustering method. Building a classifier helps us infer the class of incoming jobs and observe existing patterns and evolving trends in the system. 



For the classification tasks, we used $60$K labeled data across $119$ classes. The data is partitioned into $80\%$ and $20\%$ for training and testing, respectively. The input of the classification model is the latent feature generated from the GAN model, and the output is the label for a class between $0-118$ if the datapoint belongs to the known class otherwise it will be classified as unknown.

\begin{figure}[t]
  \centering
  \includegraphics[width=0.8\linewidth]{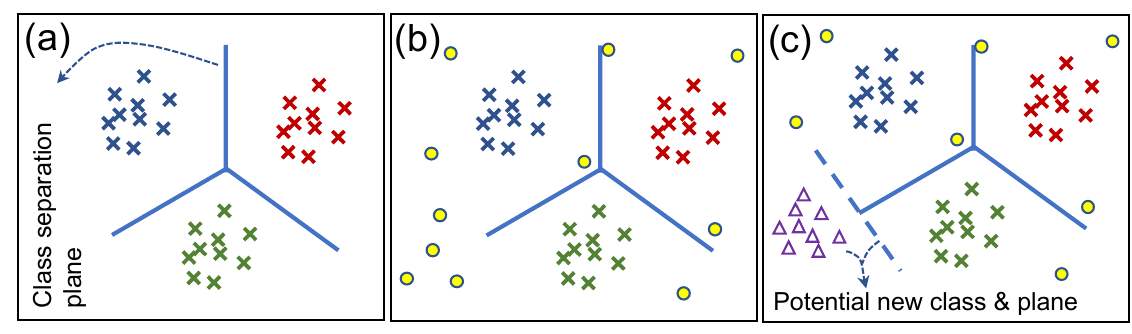}
  \caption{Illustration of open-set recognition for HPC power profile classification. (a)the dataset (represented by \texttt{x}) classified into 3 classes by a classifier. (b) datapoints belonging to these 3 classes 
  (c) new datapoints (triangle) require retraining the neural-net classifier to create a new separation plane for the new class. 
}
  \label{figs:openset-illustration}
\end{figure}

\subsubsection{Open-set Classification}
It is a classification and pattern recognition technique that is useful when the model needs to classify data not only for the already known classes but also when it encounters data that does not belong to any of the predefined classes\cite{Toward_Open_Set_Recog,miller2021class}. In contrast to closed-set, open-set classification recognizes the existence of ``unknown" or ``out-of-distribution" data. It first identifies if the data is from a known or unknown distribution, and if it is from the known set then determines the class of the data. Illustration in Figure~\ref{figs:openset-illustration} describes the different scenarios of the data and classification boundary. Figure~\ref{figs:openset-illustration}(a) illustrates the typical machine learning classification setup where all data points are classified into three classes by classification separation plane. Figure~\ref{figs:openset-illustration}(b) shows data points represented by circles which may not be classified into any of the three classes, this is a more real-world scenario and usually a challenge for the traditional machine learning classifier. 
Another scenario in an evolving real-world system is shown in Figure~\ref{figs:openset-illustration}(c). It shows that when the classification model was initially trained, there were only three classes, but over the period, a large number of data points were labeled as unknown, and potentially some of the data points can form a new class (shown by triangle points).

In prior work, several efforts have addressed open-set data for classification tasks. 
Open-set classifiers are categorized into generation-based~\cite{osr_gen1,osr_gen2} or distance-based threshold methods~\cite{Toward_Open_Set_Recog,miller2021class}. Generation-based methods generate the data
from likely unknown classes and then keep that as one of the classes during the training of the traditional classifiers.
The distance-based technique depends on the distance of the data points from the representative point of each class. 

For building an open-set classifier, we use the Class Anchor Clustering (CAC) loss function instead of cross-entropy loss for training neural networks because CAC ensures the clustering of vectors in the logit layer space for open-set classification~\cite{miller2021class}. CAC was proposed as a distance-based open-set classification method for image datasets. 
In this work, we leverage CAC for training an open-set classifier based on the timeseries latent features generated from the output of the GAN model.
CAC loss function is a combination of modified tuplet loss \cite{openset_tuplet} and anchor loss \cite{miller2021class}.
Tuplet loss (Equation \ref{eqn:tuplet}) maximizes the gap between the distance to the correct and incorrect class centers. 
\begin{equation}
    \mathcal{L}_{tuplet}(\textbf{x},y) =  log\left ( 1 + \sum_{j\ne y}^{N}\exp(d_y - d_j)  \right )
    \label{eqn:tuplet}
\end{equation}
$N$ is the number of known classes, and $d_j$ is the Euclidean distance between the center of the $j^{th}$ class and a vector of logit layer of classifier/model $f$ for data $\textbf{x}$, i.e., $f(\textbf{x})$. Anchor loss (Equation~\ref{eqn:anchor}) ensures that the absolute distance of the projection of \textbf{x} at the logit layer ($f(\textbf{x}$)) to the actual class center ($c_y$) is minimal.
\begin{equation}
    \mathcal{L}_{anchor}(\textbf{x},y) =  \left \|  f(\textbf{x}) - c_y\right \|
    \label{eqn:anchor}
\end{equation}

Thus, CAC loss is  $\mathcal{L}_{CAC}=\mathcal{L}_{tuplet}+\lambda~\mathcal{L}_{anchor}$, where $\lambda$ is a hyperparameter. 
The class center for all the known classes is calculated in the logit space based on the logit layer values for each label's data points. For a new data point, we extract its mapping in the logit space and calculate its distance from all the cluster centers. If the minimum distance from all the cluster centers exceeds the threshold distance, we label the data point as unknown. Otherwise, we assign the class label the same as the one with which the data point has the minimum distance. The CAC algorithm improves the ability of the classifier to distinguish between known and unknown classes, making it reliable and effective for open-set recognition in real-world scenarios. Although CAC classification in~\cite{miller2021class}~is applied to the image datasets, we modified our model for the tabular data for the features generated from the timeseries.

\subsection{Iterative workflow}
\label{subsec:IterativeWorkflow}
\begin{figure}[t]
  \centering
  \includegraphics[width=0.7\linewidth]{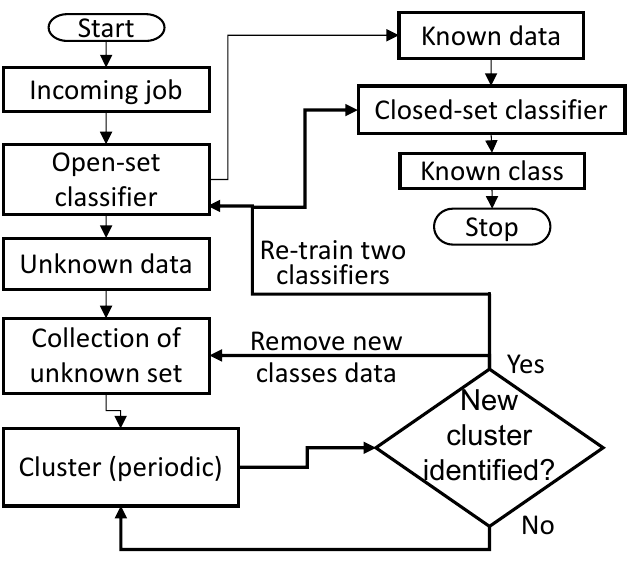}
  \caption{Iterative workflow section of the pipeline. The pipeline also explores new classes as the HPC system is continuously exposed to new jobs. It periodically runs the clustering to generate new clusters, and it is the human who decides if the new clusters are to be incorporated into the pipeline. The decision box in the flow diagram is where the human is involved.}
  \label{figs:iterate-workflow}
\end{figure}

The iterative workflow ensures that the pipeline can adapt to the evolving workload patterns. The open-set classifier describe above (in Section~\ref{sec:details:openset}) is reliable to determine the known and unknown classes based on already defined labels. However, in the real-world, HPC systems encounter jobs exhibiting new patterns. If the new patterns are too few or inconsequential we can continue to treat those as unknown data points. However, if the frequency of the jobs of new applications is frequent or consequential, our prediction pipeline should be able to capture it. Thus we designed an iterative workflow, by which we periodically (at 3-4 months intervals) apply clustering to the dataset from the unknown class. If we discover a cluster having a large number of new data points, we label them as a new class. Thus increasing the coverage of known data. Adding a new class requires us to develop a new open-set classifier to handle the new class. If the clustering method does not result in a new class then there is no need to update classification models. We repeat this process periodically and the Figure~\ref{figs:iterate-workflow} shows the iterative flowchart section from the figure in the end-to-end pipeline describing the movement of the data in more detail for periodic iteration of the pipeline. This scenario is illustrated in Figure~\ref{figs:openset-illustration}(c). Figure~\ref{figs:iterate-workflow} shows the component that needs to be updated and data flow during the iteration. The decision box is the step where a human (expert from the facility) is involved in ascertaining the new cluster.


\section{Evaluation}
\label{sec:results}


In this section, we present the analysis of classes and we also present the results of the performance of closed-set and open-set classification models.

\subsection{Analysis of Classes}
\label{subsec:results:analysis}

\begin{figure}[t]
  \centering

  \includegraphics[width=0.9\linewidth]{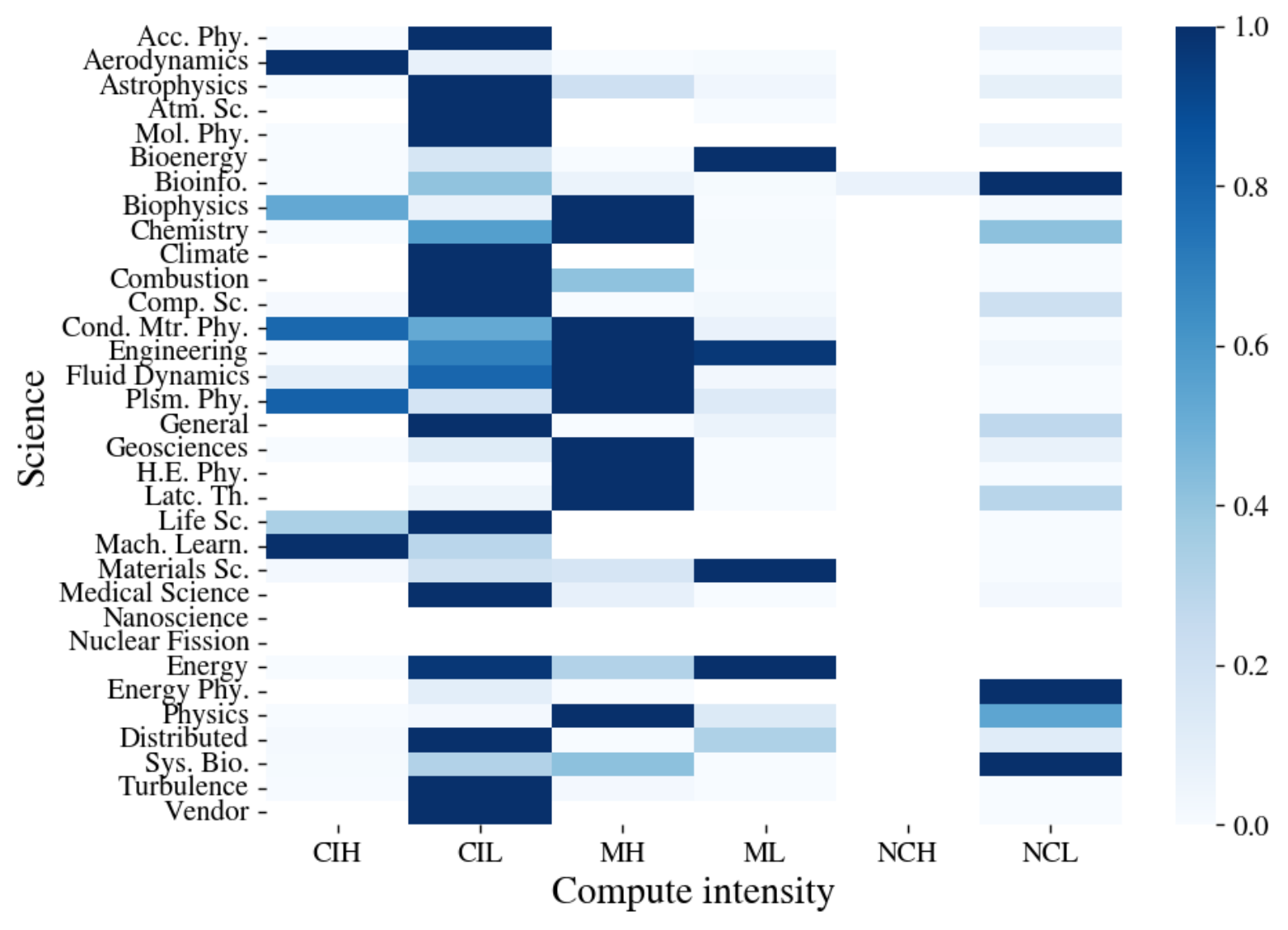}  
    \caption{Jobs distribution science-wise. The values of the heatmap are normalized row-wise to illustrate the majority of job types from each science domain. The \texttt{x-axis} labels are described in Table~\ref{tab:intensity}.}
  \label{fig:science_domain}
\end{figure}

To have a system-wide monitoring capability, we identified major and important patterns exhibited by the HPC workloads. We grouped about 60K jobs into 119 classes using the clustering method (groups are shown in Figure~\ref{fig:all_class}). 
The dataset consisted of about
$200K$ jobs that were fed to the clustering method. After applying the clustering, we got $119$ classes with about $60K$ jobs. The clustering method had more than $119$ clusters, but the other clusters were either small, i.e., less than $50$ data points, or clusters
with mixed patterns (non-homogeneous). 
For our analysis, we leveraged the $60K$ data points that belong to $119$ clusters. The high-level distribution of the workloads based on the job pattern and magnitude, along with classification labels, is shown in Table~\ref{tab:intensity}. The intensity of the background pink color in subplots of Figure~\ref{fig:all_class} shows the distribution of jobs over 119 classes. Classification of jobs based on the compute intensity provides an understanding of the nature of workloads.
Figure~\ref{fig:science_domain}~shows the domain-science wise distribution of jobs. The number of jobs is normalized between $0-1$ (from minimum to maximum) based on the number of jobs in each of the $6$ groups (shown on \texttt{x-axis}). 
The dark blue color shows the majority of jobs from the corresponding science domain contribute to that specific type of job. High intensity in the first column corresponds to \texttt{Aerodynamics}, and \texttt{Mach. Learn.} shows that most of the jobs from these domains are compute-intensive jobs with high magnitude. Likewise, several domains have high intensity in the second and third columns showing a large number of jobs from those science domains are either from compute-intensive or mixed-operation jobs. 

\begin{table}
   \footnotesize
    \caption{Intensity-based grouping. Classes column correspond to class numbers in Figure~\ref{fig:all_class}.}
    \label{tab:intensity}
    \centering
    \begin{tabular}{lcccc}
        \toprule
        Classification & Classes & Resources & Labels & Samples\\
        \midrule
        \multirow{2}{*}{Compute Intensive} & \multirow{2}{*}{0-20} & High & CIH & 6863\\
         &  & Low & CIL &8794\\
        \midrule 
        \multirow{2}{*}{Mixed-operation} & \multirow{2}{*}{21-92} & High & MH & 22852\\
         &  & Low & ML &9591\\
        \midrule 
        \multirow{2}{*}{Non-compute} & \multirow{2}{*}{92-118} & High & NCH &19\\
         &  & Low & NCL&5154\\
         \bottomrule
    \end{tabular}
\end{table}

\subsection{Closed-set Classification}
\label{subsec:results:closed_set_detection}
Closed-set classification is a traditional classification approach. The model predicts the class of incoming data points among one of the known classes and assumes that the incoming data will be from one of the known classes.
We evaluate the model's performance with the increased number of known classes. We show the accuracy results for the closed-set classification model with a varying number of known classes in Table~\ref{tab:accuracy_table}. 
As the number of classes increases, we expect a little drop in the model's performance. Performance drop reflects the increasing complexity of the problem because, with more classes, the model has to generate more class separation planes, illustrated in Figure~\ref{figs:openset-illustration}(c).

\begin{figure}[t]
  \centering

  \includegraphics[width=0.9\linewidth]{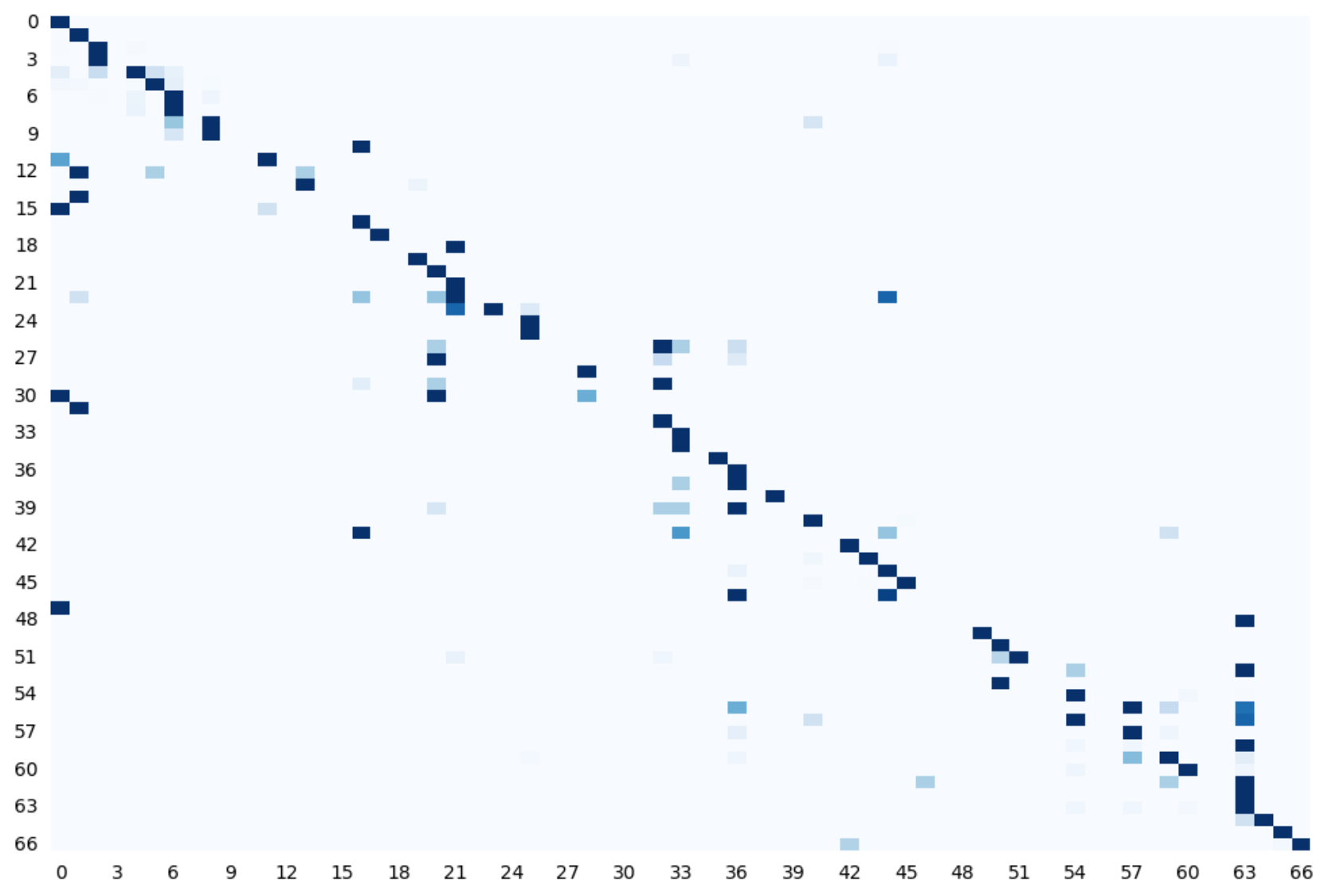}  
    \caption{Heatmap representing a class-wise performance of closed-set when
known classes are 0-66 in Table~\ref{tab:accuracy_table}. The diagonal dark regions reflect that most classes were correctly classified. While for a few other classes, the model experienced difficulties in predicting as accurately as other classes.}
  \label{fig:heatmap_known}
\end{figure}

In addition to the models' overall accuracy, it is useful to know the class-wise accuracy of the model. We generate a heatmap illustrating the class-wise model performance, as depicted in Figure~\ref{fig:heatmap_known}. The heatmap is normalized row-wise to give equal weights to all classes in the diagram, independent of the number of data points in each class. We observe that a large number of classes were correctly classified. However, we also observe that for a few classes, the model performance accuracy was relatively low, as shown by dark regions away from the diagonal. 

The reason for low accuracy performance for some classes could be due to their similarity to other classes or the model would have needed more data to train to accurately predict those classes. The high overall accuracy of the model suggests that these classes have less number of data points.

\newif\ifshowcomment

\ifshowcomment
\begin{table}[]
\caption{Accuracy of the open-set and closed-set classifier with varying number of known classes (shown in Figure~\ref{fig:all_class}).}
    \label{tab:accuracy_table}
    \centering
    \begin{tabular}{ccc}
        \toprule
           &    \multicolumn{2}{c}{Accuracy} \\
        Known classes & Closed-set & Open-set \\
        \midrule
        0-16 & 0.93 & 0.93 \\
        0-32 & 0.93 & 0.92 \\
        0-66 & 0.92 & 0.91 \\
        0-92 & 0.89 & 0.89 \\
        0-110 & 0.88 & 0.87 \\
        0-118 & 0.86 & NaN \\
        \bottomrule
    \end{tabular}
\end{table}
\else
\begin{table}
    \footnotesize
    \caption{Accuracy of the open-set and closed-set classifier with varying numbers of known classes (shown in Figure~\ref{fig:all_class}).}
    \label{tab:accuracy_table}
    \centering
    \begin{tabular}{cccccccc}
        \toprule
        Known Classes& 0-16 & 0-32 & 0-66 & 0-92 & 0-110 & 0-118 \\
        \midrule
        Closed-set & 0.93 & 0.93 & 0.92 & 0.89 & 0.88 & 0.86 \\
        Open-set & 0.93 & 0.92 & 0.91 & 0.89 & 0.87 & NaN \\
        \bottomrule
    \end{tabular}
\end{table}
\fi

\subsection{Open-set Classification}
\label{subsec:results:open_set_detection}

Open-set classification is an approach to handle incoming data from unknown classes or unknown distributions. It is critical for real-world applications to correctly predict the known classes and equally important to not predict when the incoming data is from the unknown classes.
We determine the model accuracy for correctly classifying unknown data by varying the number of known classes as shown in Table~\ref{tab:accuracy_table} and keeping the remaining classes as unknown class. 
The model consistently performs with high accuracy showing that the model reliably differentiates between known and and unknown classes. The last row has \texttt{NA} value for the open-set column because, when we make all $119$ classes from $0-118$ as known, then there are no unknown classes. 

We observe that the accuracy slightly drops as the number of known classes increases. The reason is that the known class covers more regions in the space, and the separation between classes reduces. Thus, finding the right threshold becomes more complex and results in a minor performance drop. 

\subsection{Future Classification and Iterative Workflow}
\label{subsec:results:iterative_wrokflow}

To estimate the reliability of the classification model, particularly for the real-world environment, we need to assess the classification performance on the future dataset by training on the historical data. We show the results in Table~\ref{tab:temporal_tab} for the known-set classification on the future data when the model was trained only on historical data. The first row in  Table~\ref{tab:temporal_tab} shows the average accuracy of the model when it was trained only on one month of data. The accuracy column shows the accuracy of the model when the test data is 1-week, 1-month, and 3-months in the future. Again, for the first row, if the model is trained on January data, then we calculate the prediction accuracy on the known set from 1-week of February, the complete 1-month of February, and the predictions for 3-months (February, March, and April). We repeat this process for every month until the future data is available. So when the test data is only 1-month, we train only up to November, 1-month at a time. Likewise, for 3-months, we train up to October, 3-months at a time. We report the average values of the accuracy in Table~\ref{tab:temporal_tab}. The number of known classes increases as the number of months increases. We also report the average number of the known classes seen by models when trained on 1,3,6,9 and 11 months.

The reliability of a real-world problem like ours not only depends on the classification of the future data from known distribution, but it is also critically important to know which future data does not belong to any of the already known classes. We test the performance of the open-set classification model when it is encountered with handling future datasets that belong to unknown classes. The unknown set part in Table~\ref{tab:temporal_tab} shows the accuracy of the open-set classification when trained for only 1,3,6,9 and 11. 

In Table~\ref{tab:temporal_tab}, we also observe that as the number of months increases, the number of known classes increases. It is the reflection of the nature of the HPC system. The workflow pipeline open-set and closed-set can classify known classes and unknown classes respectively, with a high accuracy. The result shows that the pipeline reliably accommodates the evolving workloads on HPC.

\begin{table}[]
\caption{Accuracy of the open-set classifier for future time points up to 1-week, 1-month, and 3-months based on the training dataset of 1,3,6,9, and 11 months.}
\label{tab:temporal_tab}
\footnotesize
\begin{tabular}{lccccc}
    \toprule
     \multirow{2}{*}{Set} & \multirow{2}{*}{\begin{tabular}[c]{@{}l@{}}Trained data \\ (in months)\end{tabular}} & \multirow{2}{*}{\begin{tabular}[c]{@{}l@{}}Known \\ classes\end{tabular}} & \multicolumn{3}{c}{Accuracy (on future data)} \\
     &  &  & 1-week & 1-month & 3-months \\
         \midrule
    {\multirow{5}{*}{\rotatebox[origin=b]{90}{(a) Closed-set}}} & 1 & 52 & 0.76 & 0.71 & 0.66 \\
     & 3 & 80 & 0.79 & 0.81 & 0.66 \\
     & 6 & 96 & 0.90 & 0.82 & 0.64 \\
     & 9 & 96 & 0.87 & 0.92 & 0.49 \\
     & 11 & 118 & 0.76 & 0.58 & X\\
     \midrule
    {\multirow{5}{*}{\rotatebox[origin=b]{90}{(b) Open-set}}} & 1 & 52 & 0.91 & 0.91 & 0.90 \\
     & 3 & 80 & 0.87 & 0.86 & 0.85 \\
     & 6 & 96 & 0.90 & 0.89 & 0.89 \\
     & 9 & 96 & 0.85 & 0.84 & 0.82 \\
     & 11 & 118 & NA & 0.85 & X\\     
     \bottomrule
\end{tabular}
\end{table}

\subsection{Varying Threshold Distance for Open-set Model}
\label{subsec:results:open_set_distance}
\begin{figure}[t]
  \centering

  \includegraphics[width=0.9\linewidth]{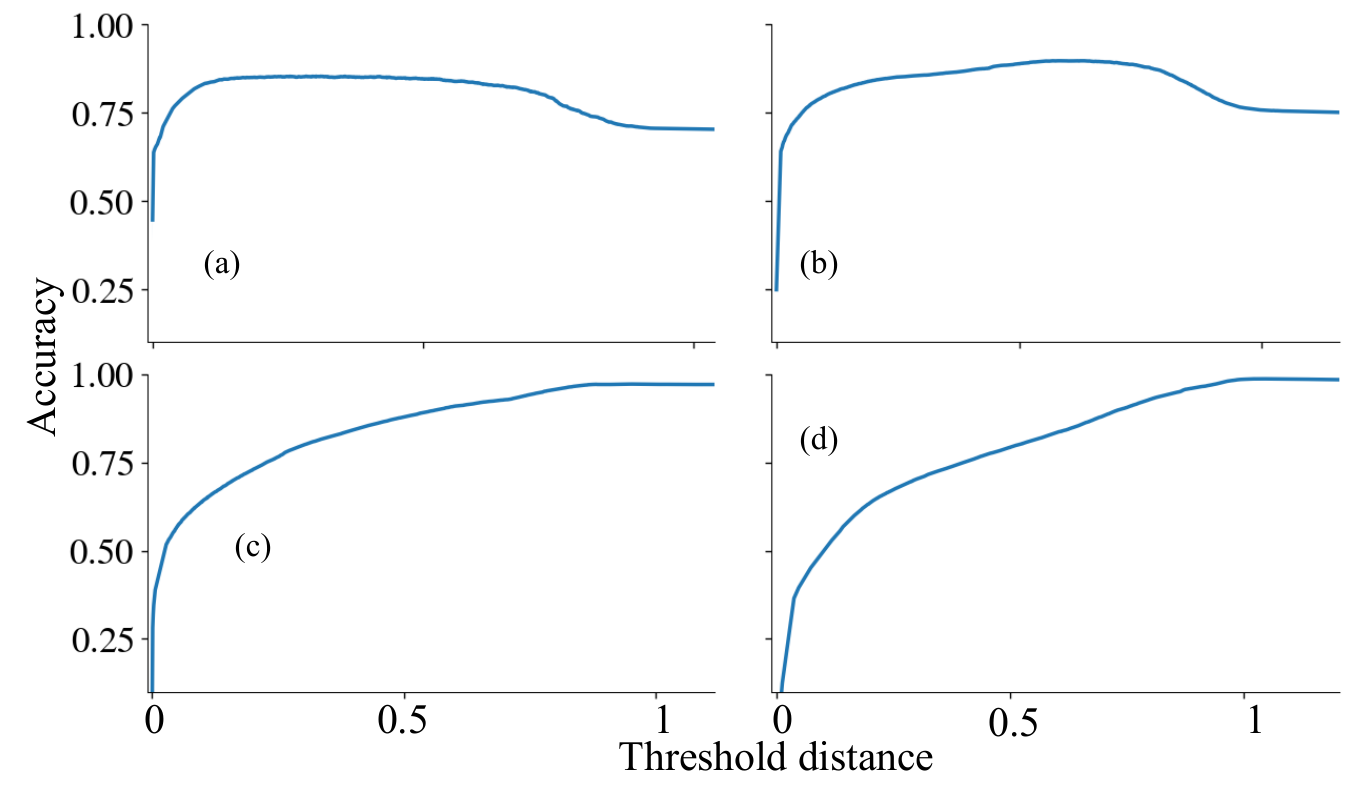}  
    \caption{The plots show the accuracy of the open-set classifier based on the threshold distance from the cluster centers. The four plots (a)-(d) correspond to the known data from four rows 1,3,6, and 9 in Table~\ref{tab:temporal_tab}. The data from remaining classes were considered unknown for evaluation.  
    }
  \label{fig:threshold}
\end{figure}

The open-set classification model differentiates the unknown datasets from the known data by calculating the distance of the data points from the cluster centers. If the distance of the data point is less than the threshold value, we classify the data point as from the known class. Otherwise, we classify it as from the unknown class. We calculate the center of the clusters from the already labeled data during training at the penultimate layer in the classification model. For new data, we extract the feature values at the logit layer and calculate their distance from the cluster centers. 

We illustrate the effect of threshold distance on the performance accuracy in Figure~\ref{fig:threshold}. The four subplots correspond to the data from Table~\ref{tab:temporal_tab}. We select the known classes corresponding to the months from one to nine, and the remaining classes are considered unknown for the experiment. The \texttt{y-axis} represents the accuracy, and the \texttt{x-axis} represents the normalized value of threshold distance from the cluster centers. We observe that when the threshold distance is small, the model has poor accuracy, and as we increase the distance, the model accuracy increases. Towards the higher threshold distances, model accuracy starts to drop. It demonstrates that finding the correct threshold value is also essential for optimal accuracy. 


\vspace{-2pt}
\section{Related Works}


\subsubsection*{\textbf{Data driven job classification}}
In recent years, there has been an increasing focus on utilizing data-driven methods to classify jobs in HPC systems \cite{bang2020hpc,fan2021dras}, which rely on machine learning techniques to evaluate the characteristics of jobs and predict job behavior (i.e. job duration \cite{naghshnejad2018adaptive,galleguillos2018data,wyatt2018prionn,mckenna2016machine}, job resource usage \cite{bose2021hpcgcn,sirbu2016power,matsunaga2010use}, job scheduling \cite{galleguillos2020accasim,klusavcek2020alea}). \cite{presser2022towards} presents the feasibility of real-time classification of HPC workloads, where the focus is given to the detection of malicious HPC workloads, but the work lacks in having the experimental dataset, as malicious workloads were missing to perform their classification evaluation. The authors address the job classification problem \cite{tsujita2021job} of predicting two classes the high-power-consuming jobs and low-power-consuming jobs by using correlation analysis of job activities in CPU, memory, file I/O, and compute node layouts using supervised classification models. In contrast, we focus on clustering jobs based on the jobs' power consumption and learn about different patterns the jobs exhibit, which is useful for predicting the power consumption pattern of the incoming jobs. This work will also assist the HPC power facility with improved power awareness of the workloads.

\subsubsection*{\textbf{Timeseries clustering methods}}
Variable length timeseries clustering is an active area of research, and several methods have been proposed in recent years, differing in their distance measures \cite{muller2007dynamic,gharghabi2018matrix}, clustering algorithms, and applications \cite{amaral2022summertime}. There are two categories for timeseries clustering algorithms \cite{aghabozorgi2015time,liao2005clustering,alqahtani2021deep,javed2020benchmark}: raw-data-based or shape-based \cite{begum2015accelerating,paparrizos2017fast,wang2019hierarchical}, and feature-based \cite{tiano2021featts,tavenard2020tslearn,fulcher2017feature}, and our work falls into the latter one. \cite{trosten2019recurrent} presents a joint clustering and feature learning framework based on deep learning, where a recurrent network is trained to embed, each timeseries in a vector space such that a divergence-based clustering loss function can discover the underlying cluster structure, but it cannot handle timeseries with longer lengths due to the computational complexity of training the deep neural network. Some of the recent works \cite{madiraju2018deep, ma2019learning} are in the initial stages. However, none of these works have shown experimental analysis with HPC operational data.   

\section{Conclusion}

In this work, we have presented the design and implementation of a system-wide HPC monitoring system based on their power profile for a (pre-)exascale leadership-scale system. The end-to-end pipeline provides contextualized labels for job-level timeseries and then leverages them for building an open-set classification model. We introduced a set of features calculated from the timeseries signal, which have proven to be significant in classifying HPC job power profiles based on the intensity and patterns, and help us understand the most frequent and critical patterns encountered by the HPC system (covered in Figure~\ref{fig:all_class} \& Table~\ref{tab:intensity}). 

We grouped about $60K$ jobs into $119$ clusters. Each cluster represents a unique pattern such as frequency of swings, magnitude of swings, and rising and falling slopes, and reflects the behavior of underlying workloads. This enabled us to have a system-wide overview at job-level as well as the system-level. 
We analyzed jobs' power patterns based on the science domain and identified the most dominant power patterns exhibited by jobs of each science domain. 
For continuous monitoring, the classifiers under closed-set and open-set scenarios resulted in high accuracy under different scenarios. 

To emulate the real-world setup, we also tested the classification models by training on historical data and then making
predictions on the future data. Both open-set and closed-set models performed with significantly high accuracy values. 
In the future, we aim to enhance the performance of the model across all classes.
We intend to improve the clustering technique to find new clusters more effectively. Furthermore, our goal is to attain complete automation, by removing manual visualization of clusters during iterative step, particularly when adding a new class to the classification model.



\bibliographystyle{IEEEtran}

\end{document}